\documentclass[letterpaper,12pt]{article} 

\usepackage{opticameet3} 

\newcommand\authormark[1]{\textsuperscript{#1}}

\usepackage{amsmath,amssymb}
\usepackage[colorlinks=true,bookmarks=false,citecolor=blue,urlcolor=blue]{hyperref} 

\begin{document}

\title{Utilizing Quantum Fingerprints in Plant Cells to Evaluate Plant productivity}


\author{Umadini Ranasinghe,\authormark{1*} Abigail L. Stressinger,\authormark{2*} Guangpeng Xu,\authormark{1} Yasmin Sarhan,\authormark{1} Fred Harrington,\authormark{3} James Berry,\authormark{2} Tim Thomay\authormark{1}}

\address{\authormark{1} Department of Physics, State University of New York at Buffalo, New York

\authormark{2} Department of Biology, State University of New York at Buffalo, New York

\authormark{3} Helios-NRG, LLC, East Amherst, New York}

Overcoming the strong chlorophyll background poses a significant challenge for measuring and optimizing plant growth. This research investigates the novel application of specialized quantum light emitters introduced into intact leaves of tobacco \textit{(Nicotiana tabacum)}, a well-characterized model plant system for studies of plant health and productivity. Leaves were harvested from plants cultivated under two distinct conditions: low light (LL), representing unhealthy leaves with reduced photosynthesis. and high light (HL), representing healthy leaves with highly active photosynthesis. Higher-order correlation data were collected and analyzed using machine learning (ML) techniques, specifically a Convolutional Neural Network (CNN), to classify the photon emitter states. This CNN efficiently identified unique patterns and created distinct fingerprints for \textit{Nicotiana} leaves grown under LL and HL, demonstrating significantly different quantum profiles between the two conditions. These quantum fingerprints serve as a foundation for a novel unified analysis of plant growth parameters associated with different photosynthetic states. By employing CNN, the emitter profiles were able to reproducibly classify the leaves as healthy or unhealthy. This model achieved high probability values for each classification, confirming its accuracy and reliability. The findings of this study pave the way for broader applications, including the application of advanced quantum and machine learning technologies in plant health monitoring systems.


\textbf{Abbreviations: Quantum dots (QDs), Low light (LL), High light (HL), Machine learning (ML), Convolution neutral network (CNN), Green fluorescent protein (GFP), Hanbury-Brown and Twiss (HBT), Beam splitter (BS), Avalanche photodiodes (APDs), Correlation board (CB), photon correlation spectroscopy (PCS), Photosynthetically active radiation (PAR), Photoluminescence (PL)}. 

\section{Introduction}

Autofluorescence, the natural light emission from plant tissues, poses a significant obstacle in accurately studying their fluorescent properties, sometimes leading to inaccurate determinations of photosynthetic activity \cite{Murchie2013Chlorophyll,Maxwell2000Chlorophyll,Lawson2024Imaging}. This phenomenon, primarily driven by chlorophyll emissions and other light gathering proteins \cite{Magdaong2018Photoprotective}, can be affected by multiple variable factors, including quenching and overlapping signals, making it difficult to quantify and standardize the multiple parameters associated with overall photosynthetic health and productivity.  Quantum dots (QDs), with their unique property of emitting only one photon at a time, help to overcome this strong chlorophyll fluorescence commonly found in plants.

Due to their distinct of light, QDs as quantum light emitters are highly specific and reflective of their surrounding conditions. Therefore, quantum light from QD-transformed species is reflective of the environmental and intracellular conditions of the organism. This property has led to the use of quantum dots in plant systems \cite{Glauber1963Quantum}, primarily as enhancers of productivity and biosensors for pathogen and pollutant detection \cite{Liu2009Antibacterial}. Furthermore, QDs have been utilized to monitor nutrient levels in plants, providing critical insights into their nutritional status and enabling nutrient management strategies \cite{Ghormade2011Perspectives, Khot2012Applications, Liu2015Potentials}. However, most recent applications have focused on utilizing QDs to mitigate abiotic stressors and enhance plant growth \cite{Chen2024Recent, Khan2024Nanoparticles, Oyebamiji2024Recent, Tripathi2022Crosstalk, Yan2024SiO2,Liu2009Antibacterial}. As quantum light emitters, QDs have the ability to monitor plant health and distinguish sub-optimal growth conditions. Yet, minimal research has been conducted on the role of QDs as evaluators of plant productivity.

Understanding how QDs are used as quantum light emitters requires first recognizing how the concept of light as discrete energy quanta has led to advancements in studying photon number properties. These developments have established a foundational concept central to our current quantum optics experiments. In an early 1956 study, Hanbury Brown and Twiss \cite{BROWN1956Correlation} conducted a groundbreaking experiment that distinguished between thermal and Poissonian light, marking a pivotal moment in the study of optical coherence. This foundation was further established in a 1963 study by Glauber and Sudarshan \cite{Glauber1963Photon} introduced the quantum theory of photon correlations. Their work focused on higher-order factorial moments of photon-number distributions, providing a framework to analyze and understand the behavior of light at the quantum level \cite{Glauber1963Photon,Glauber1963Quantum,Sudarshan1963Equivalence,Laiho2022Measuring}. Their findings set the stage for advancements in understanding the properties of light sources, bridging classical and quantum optics.

The exploration of higher-order photon correlations has since led to numerous applications, particularly in quantum light emitters. \cite{Powers2023,Pusey2011On,Powers2023event}. Furthermore, this technology can be used in a range of biological applications \cite{Liu2024Generalized,Nicora2020Integrated}.Most notably, photon correlation techniques have been applied in cancer diagnostics. Detecting and treating cancer remains a major challenge in medicine. Optics and photonics technologies have applied principles of physics to enhance diagnostic methods \cite{2020Multimodal}. Also, Photon correlation spectroscopy (PCS), a specialized technique within photon correlation that analyzes fluctuations in scattered light to determine the size and motion of particles in a sample, has been used to study proteins and other biomacromolecules in aqueous solutions. PCS has been widely applied to investigate conformational changes in proteins, their aggregation, and interactions with other molecules. These studies help to reveal important native functions of proteins, DNA, RNA, and even microorganisms like \textit{Escherichia coli}, \textit{Pseudomonas putida}, and \textit{Dunaliella viridis}
\cite{Gun2003Photon,Steer1985Laser}. Furthermore, time correlated single photon counting (TCSPC) is a technique that detects single photons from a periodic light signal, records their detection times, and creates a distribution of photons over the signal's time period. This technique has numerous biomedical applications, such as time-domain optical tomography, studying transient phenomena in biological systems, spectrally resolved fluorescence lifetime imaging, Fluorescence resonance energy transfer (FRET) experiments in live cells, and analyzing dye-protein complexes using fluorescence correlation spectroscopy \cite{Becker2004Advanced,Kress2003Time,Zachariasse1985Einzelphotonenzaehlung}. However, there has been limited research exploring photon correlation in plants.

In the past decade, data-driven approaches like Machine Learning (ML) have opened up new opportunities for quantum photonics experiments \cite{Yao2019Intelligent,Kudyshev2020Machine,Zhou2019Emerging,Virtanen2020SciPy}. ML models, known for handling large and sparse datasets, have achieved significant speedups in certain quantum measurements \cite{Hegde2020Deep,Freire2023Artificial,Liu2018Training,Yao2019Intelligent} and offer a way to overcome the limitations of traditional fitting methods, especially in the low-photon flux regime \cite{Wu2023Classification,Gao2018Experimental,Li2021Fast,Emani2021Quantum}. One notable advancement is the development of a Convolutional Neural Network (CNN)-based algorithm tailored for the rapid classification of single-photon emitters within the nitrogen-vacancy (NV) center of nanodiamonds \cite{Kudyshev2020Rapid,Dunjko2017Machine,Melnikov2018Active,Cong2019Quantum}. The CNN model improves accuracy by identifying subtle features extracted from sparse correlation data.

In this study, we aim to develop an innovative technique to optimize plant productivity using a novel quantum-based fingerprinting concept. The approach involves introducing biocompatible quantum dots (QDs) as quantum light emitters into the leaf cells of tobacco plants. Plants were grown under two distinct environmental conditions, low light (LL) and high light (HL), to assess the technique’s applicability across different growth scenarios. We hypothesize that QDs as quantum light emitters provide a significant advantage over methods using classical chlorophyll fluorescence detection, which are often masked by plant pigments and light-absorbing/quenching components in photosynthetically-active plant tissues \cite{Maxwell2000Chlorophyll, Lawson2024Imaging, Magdaong2018Photoprotective, Berry2013Photosynthetic}. Quantum light sources emit single photons with unique quantum properties, allowing them to be optically distinguished from classical multiphoton light sources.  This optical quantum differentiation is achieved by leveraging higher-order correlation functions to analyze and confirm the quantum nature of the emitted light. To interpret the data, we classify the quantum signals based on time resolved correlation patterns using a Convolutional Neural Network (CNN) model. The CNN is trained to extract subtle and complex features from the data, enabling it to accurately differentiate between the quantum properties of light that are emitted from healthy and unhealthy leaves. This is done without interference from any other non-quantum light emissions within the experimental leaf samples. This approach results in the creation of distinct quantum fingerprints for each leaf, representing their unique quantum fluorescence characteristics. Furthermore, we show that our pre-trained machine learning model can classify the probabilities of emissions from experimental categories, thereby predicting the derived quantum fingerprints as being from healthy or unhealthy leaves. The results show that quantum-based fingerprinting provides a novel and effective way to characterize optimal plant growth and monitor plant health under different conditions.

\section{Method}
\subsection{Preparation of leaves and analysis of photosynthesis}

Tobacco (\textit{Nicotiana tabacum} var. SR1) leaves were prepared under low-light (LL) conditions and high-light (HL) conditions. All growing conditions were standardized in growth chambers (24$^\circ$C, 12 hr light/dark cycle, daily watering) except for light intensity, which differed by 30× between groups. The high-light group was exposed to 430 $\mu$mol photons m$^{-2}$ s$^{-1}$ (\( 430 \times 10^{-6} \, \text{Einstein} \, \text{m}^{-2} \, \text{s}^{-1} \)), and the low-light group received only 15 $\mu$mol photons m$^{-2}$ s$^{-1}$ 
\(\left( 15 \times 10^{-6} \, \text{Einstein} \, \text{m}^{-2} \, \text{s}^{-1} \right)\) of light. 
Normal growth light conditions for tobacco plants are between 400-450 $\mu$mol m$^{-2}$ s$^{-1}$ 
\(\left( 400-450 \times 10^{-6} \, \text{Einstein} \, \text{m}^{-2} \, \text{s}^{-1} \right)\)
\cite{Biswal2012Light}. LL conditions significantly decrease rates of photosynthesis, leading to reductions in growth, stomatal conductance, intercellular carbon dioxide levels, and transpiration rates \cite{Yang2017Effects,Andersen1973Chemical}. As a result, the LL group is considered unhealthy compared to the HL group, which is expected to affect quantum measurements.

To determine the photosynthetic efficiency of each experimental group, we used the PhotosynQ MultispeQ V 2.0 fluorimeter device (Photosynq Inc. East Lansing, MI 48823 USA).  The following parameters were analyzed for each tobacco leaf: photosynthetically active radiation (PAR), non-photochemical quenching, relative chlorophyll, and photosystem I active centers. PAR was calculated by measuring the fraction of incoming light that is active in promoting photosynthesis \cite{Kuhlgert2016MultispeQ}. Non-photochemical quenching infers plant health by measuring the plant’s ability to dissipate excess light energy harmlessly as heat and is calculated by providing pulse-amplitude modulation fluorescence. The chlorophyll content of each leaf was calculated by measuring the ratio between the absorbance of red (650 nm) and infrared (940 nm) light. Finally, the number of active Photosystem I centers was determined by absorbance-based measurements (810-940 nm). 

These parameters provide a quantitative measurement of the overall photosynthetic efficiency of each experimental group of tobacco leaves. A total of sixteen leaves, eight from each experimental group, were measured for this study. 

\subsection{Biolistic transformation of leaves}

Tobacco leaves were transformed using the Biolistic PDS-1000/He Particle Delivery System, commonly known as the ``Gene Gun'' (Bio-Rad PDS-1000/He). This system was used to deliver custom gold microprojectiles coated with DNA and quantum dots onto the abaxial surface of each tobacco leaf. We modified previously published protocols for transformation of tobacco plants \cite{Staub1992Long,Farkas2009Development}. To prepare the microprojectiles, we added 0.1 M Biotin-PEG-SH-Thiol (Nanocs MW 5000) to 20~$\mu$l of 66 nm gold particles (Bio-Rad), creating a conjugated gold-biotin particle. Green fluorescent protein (GFP) report gene was delivered via the pBI121-GFP vector of a constitutive CaMV35S promoter \cite{Chen2003}. The GFP plasmid was added in a ratio of 10 $\mu$g DNA to 1.2 mg of gold-biotin particles. The particles were combined with 25~$\mu$l of 2.5 M CaCl$_2$, 10~$\mu$l of 0.1 M spermidine free base, then vortexed for 15 seconds. The DNA-coated microprojectiles were centrifuged at maximum speed for 5 seconds, washed with 100\% EtOH, and resuspended in 24~$\mu$l 100\% EtOH. Finally, 5~$\mu$l of Streptavidin Conjugate CdSe/ZnS core/shell Quantum Dots (585 nm, Invitrogen Thermo Fisher Scientific) were added to the particle solution. This linked the quantum dots to the gold microprojectile via the Biotin-PEG-SH-Thiol linker as shown Figure 1.

After being detached from their host plant, each tobacco leaf was bombarded twice at 900 psi by 8~$\mu$l of microprojectiles. Stopping screens were placed 6 cm away from the petri dish containing the leaf. After bombardment, leaves recovered in the dark for 24 hours, then were moved back to their initial growth chambers prior to examining for GFP expression and quantum fluorescence.

\begin{figure}[htbp]
\centering
\includegraphics[width=0.8\textwidth]{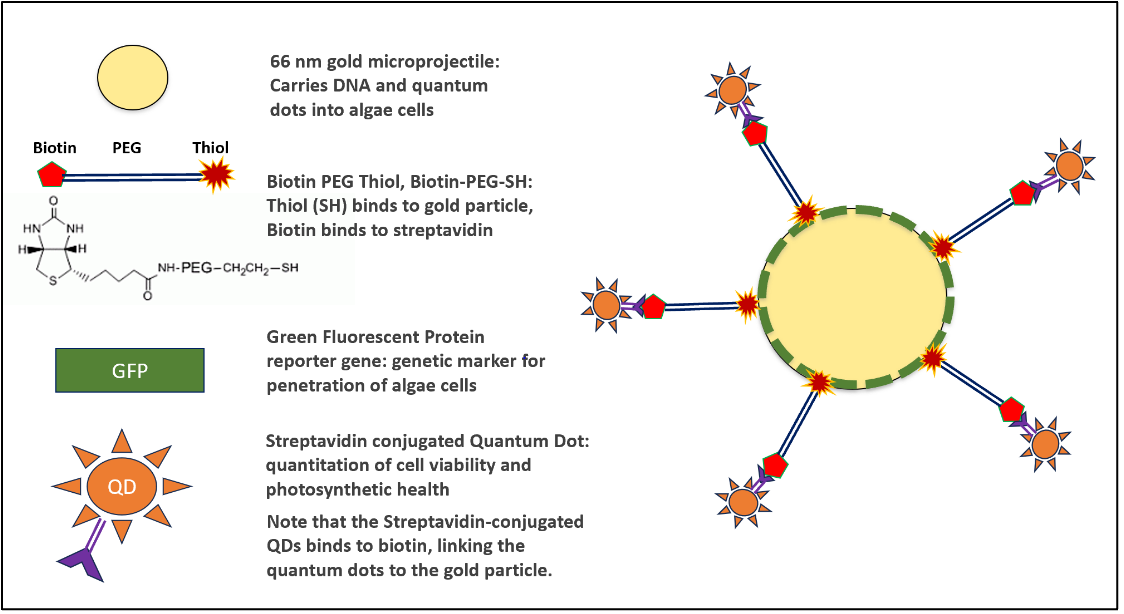} 
\caption{\label{fig:1} \textbf{Design of multi-component gold projectile designed to carry DNA and QDs into plant cells.} Biotin PEG Thiol serves as a linker between gold microparticle and QDs. Green Fluorescent Protein (GFP) serves as a visible genetic marker for successful cellular uptake, which, when expressed, fluoresces green under ultraviolet light. Streptavidin-conjugated QDs (emission spectrum of 585 nm) fluoresce orange under ultraviolet light.}
\end{figure}

\subsection{Development of quantum fingerprint}

The photoluminescence (PL) of leaves was measured to confirm the emission of QDs. A Hanbury-Brown and Twiss (HBT) setup with four detectors is shown in Figure 2. 
It was used to measure second-order correlation that describes the probability of detecting two photons at different times, providing information about the photon emission statistics. This work expands upon previous findings \cite{Thomay2017Simultaneous}, where an HBT setup is employed to study photon correlation properties. Each path leads to an Avalanche Photodiode (APD). Every photon that arrives at the APD triggers a click recording the detection time. 

\begin{figure}[htbp]
\centering
\includegraphics[width=12.0cm]{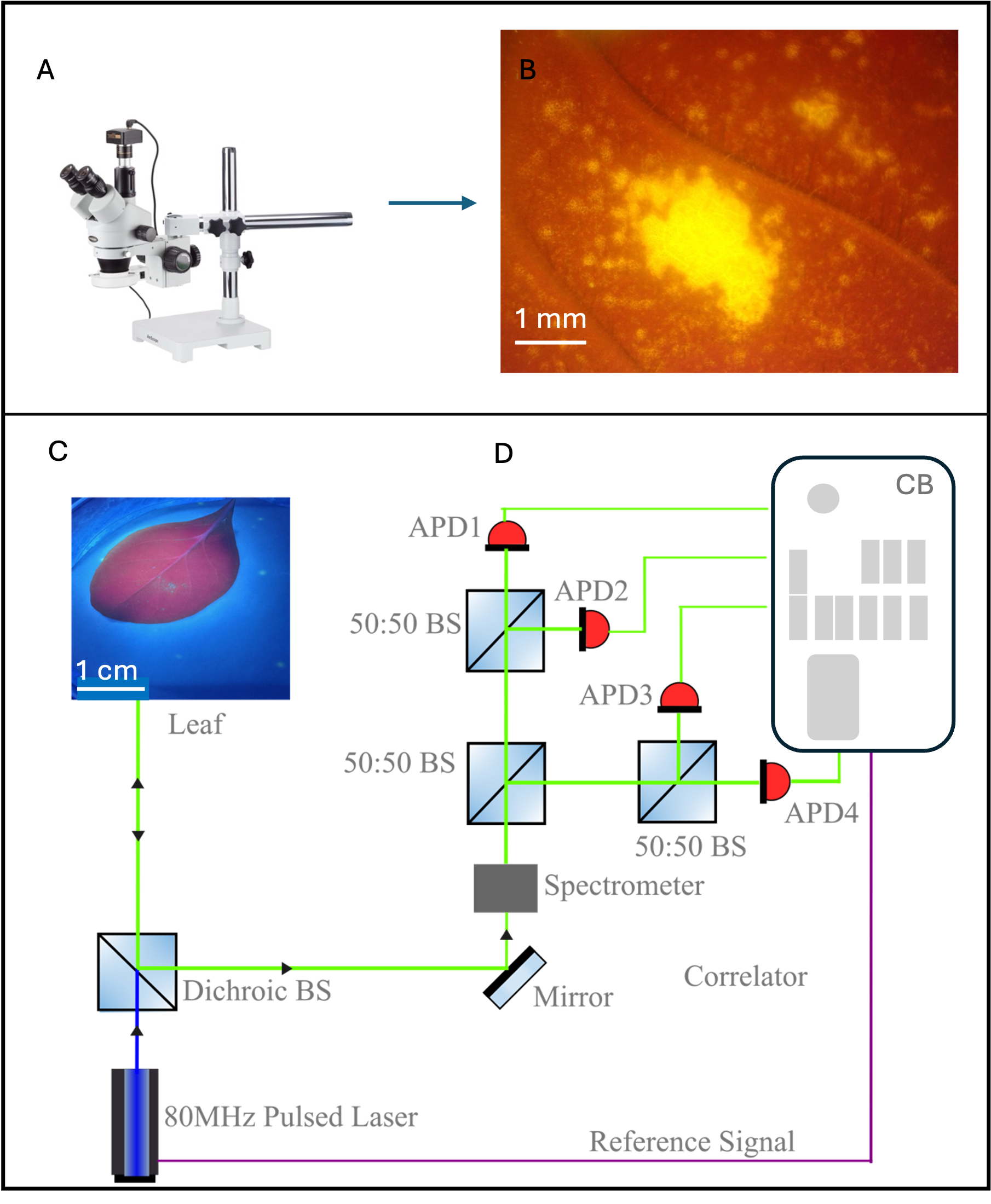} 
\caption{\label{fig:1} \textbf{Experimental layout for measuring higher-order correlation.} (a) The leaf under 365 nm UV light (b) The stereo microscope (c) The leaf under stereo microscope (d) The Hanbury Brown and Twiss (HBT) set up with 50:50 Beam Splitters:BS, and detected by Avalanche Photodiodes:APDs connected to a Correlation Board:CB.}
\end{figure}

Higher-order correlation data were collected and analyzed using a machine learning model to classify the photon states. Further details on the computational method can be found in \cite{Xu2024Optimized}. The correlation data was compiled into a 3D matrix with dimensions $(4, 4, 2x + 1)$, where $x$ is the bin number. Starting with $x = 6$, the resulting matrix has dimensions $(4, 4, 13)$, representing the $g^{(2)}$ correlation matrix.
 
 This matrix forms the foundation for creating a 2D quantum fingerprint in tobacco leaf cells by normalizing the matrix.

\subsection{Classification of probabilities}

The CNN model was employed to classify leaves based on LL or HL growth conditions. The dataset is divided into two portions, 80\% is allocated for training the model, while the remaining 20\% is reserved for testing. This division ensures that the model has sufficient data to learn patterns, relationships, and features during the training phase. The model was trained over 20 epochs, with each epoch representing a full pass through the training data. After training, the model was used to classify leaves by predicting probabilities for each sample as either healthy or unhealthy, providing a clear indication of its confidence in each classification. To further validate its performance, the trained model was tested on new, unseen data, where it successfully assigned probabilities to each sample in the validation set, determining whether a leaf is healthy or unhealthy. The model architecture and layer operations are illustrated in Figure 3. 

\begin{figure}[htbp]
\centering
\includegraphics[width=12.0cm]{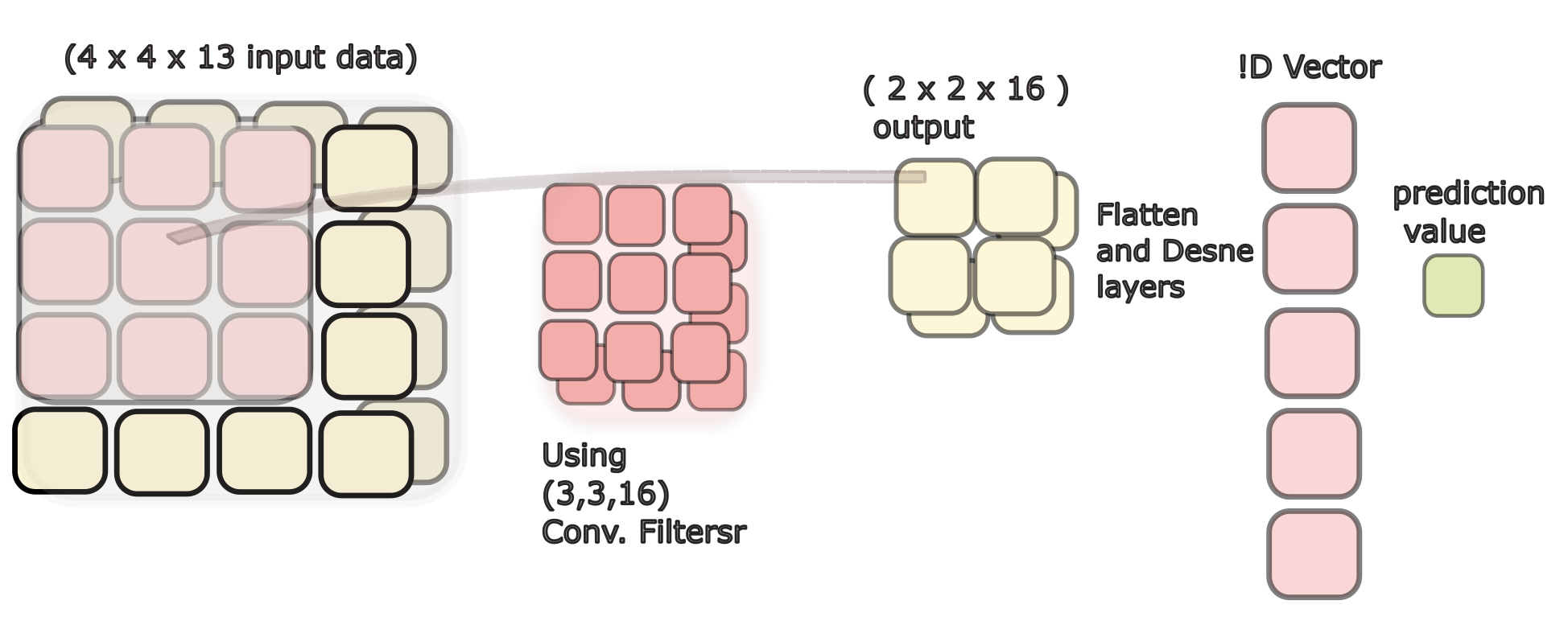} 
\caption{\label{fig:3} \textbf{Schematic of the CNN model for probability classification.} The architecture consists of a convolutional layer, followed by a Conv2D layer with 16 filters and a (3×3) kernel, activated using ReLU. A flattening layer converts the convolutional output into a 1D feature vector. The model includes dense layers for binary classification: one with 32 neurons and ReLU activation to learn complex feature representations, and a final dense layer with a single neuron and sigmoid activation for classification.}
\end{figure}

\section{Results}

This study combines quantum technology and machine learning to analyze plant health productivity. Quantum dots (QDs) were introduced into tobacco leaves to study their quantum properties under different lighting conditions. The data revealed clear differences between healthy and unhealthy leaves, visualized as unique quantum fingerprints. The CNN model accurately classified the leaves, achieving high probabilities for both healthy and unhealthy categories, demonstrating the effectiveness of this approach in plant health productivity.

To determine the productivity of tobacco leaves on a cellular level, we measured photosynthetic parameters based on light-driven fluorescence and absorbance changes via LED light. As expected, the tobacco plants used in this experiment, which were grown under reduced light intensity, exhibited lower rates of photosynthesis. Thus, the parameters for tobacco grown under HL will have high photosynthetic activity than plants grown under LL. Overall, we observed statistically significant differences in all parameters between HL and LL tobacco leaves (Fig 4). Therefore, tobacco leaves from both experimental groups have contrasting rates of photosynthesis, which will be reflected in their quantum fingerprint.

\begin{figure}[htbp]
\centering
\includegraphics[width=15.0cm]{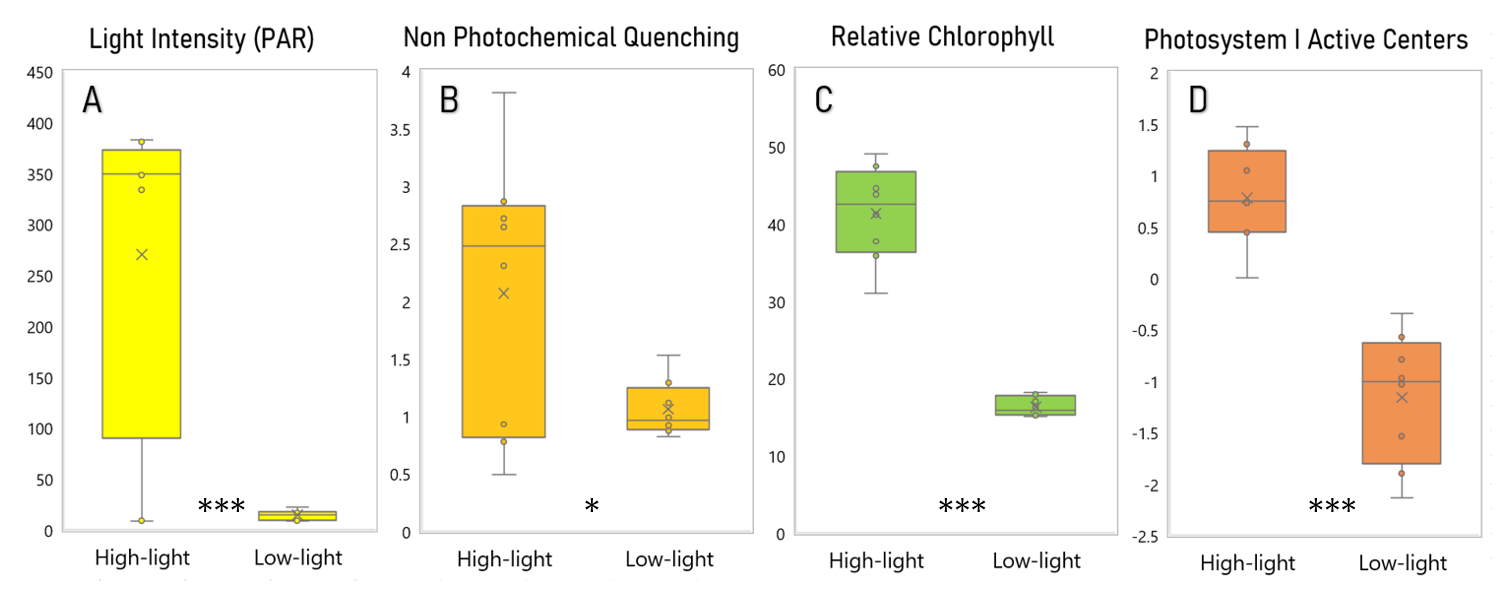} 
\caption{\label{fig:3} \textbf{Fluorescence and absorbance-based photosynthetic parameters indicate higher photosynthetic activity in HL vs. LL tobacco leaves.} (A) Light intensity via photosynthetically active radiation ($\mu$mol photons m$^{-2}$ s$^{-1}$). (B) Non-Photochemical Quenching measures each leaf’s ability to dissipate excess absorbed light energy as heat. (C) Relative chlorophyll content ($\mu$mol of chlorophyll per m$^2$ of leaf), calculated by measuring the absorbance at 650 and 940 nm. (D) The fraction of Photosystem I centers that are active is calculated via a ratio of $F_o$ (fluorescence level of a dark-adapted leaf with all Photosystem acceptors fully oxidized) to $F_m$ (the maximum fluorescence achieved when all acceptors are fully reduced). This ratio is typically between 0.75 and 0.85 for healthy/light-sufficient leaves. Statistical significance is determined by a two-sample equal variance t-test: (*$p < 0.05$), (**$p < 0.01$), (***$p < 0.001$).}
\end{figure}

\begin{figure}[htbp]
\centering
\includegraphics[width=15.0cm]{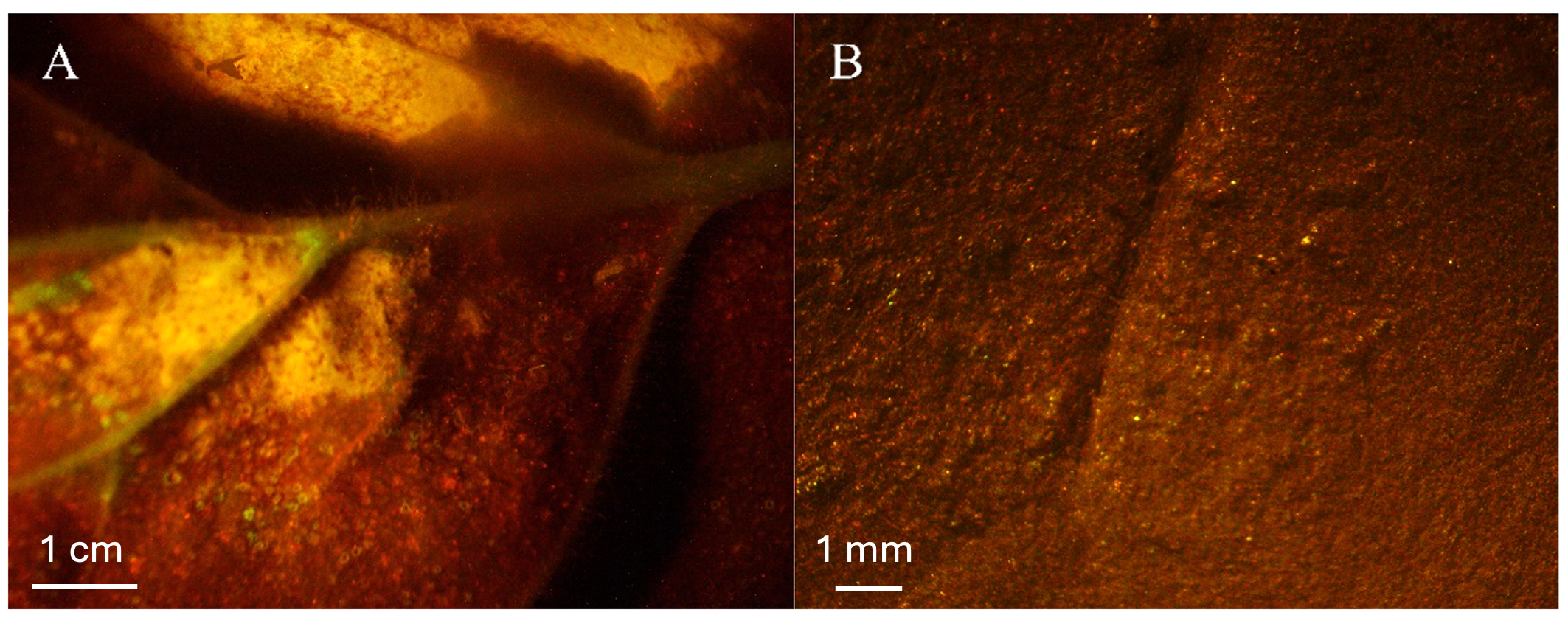} 
\caption{\label{fig:1}\textbf{Successful delivery of a multi-component microprojectile into tobacco leaf cells facilitated introduction of quantum dots.} (A) Stereoscope image of a HL tobacco leaf under ultraviolet light (365 nm) shows strong GFP and quantum dot fluorescence. (B) Stereoscope image of a LL tobacco leaf under ultraviolet light (365 nm) also shows GFP and quantum dot fluorescence, confirming successful microprojectile penetration.}
\end{figure}

The gold microprojectiles were specifically engineered to serve as carriers for both DNA and QDs into the tobacco plant cells. This design successfully allowed for the microcarrier and its components to gain cellular entry through biolistic transformation. Our constructed microprojectile links biocompatible quantum dots and Green Fluorescent Protein (GFP) reporter gene to gold nanoparticles as shown in Figure 1. Therefore, the GFP marker serves as a confirmation of delivery of the microprojectile into tobacco leaf cells. When the GFP reporter gene entered the cell nucleus, green fluorescence was observed under ultraviolet light, indicating the entry of all components, including streptavidin-conjugated quantum dots (Fig 5). Additionally, orange fluorescence was observed on both HL and LL leaves, confirming the presence of quantum dots. Both green and orange fluorescence was observed in HL and LL leaves.

\begin{figure}[htbp]
\centering
\includegraphics[width=13.0cm]{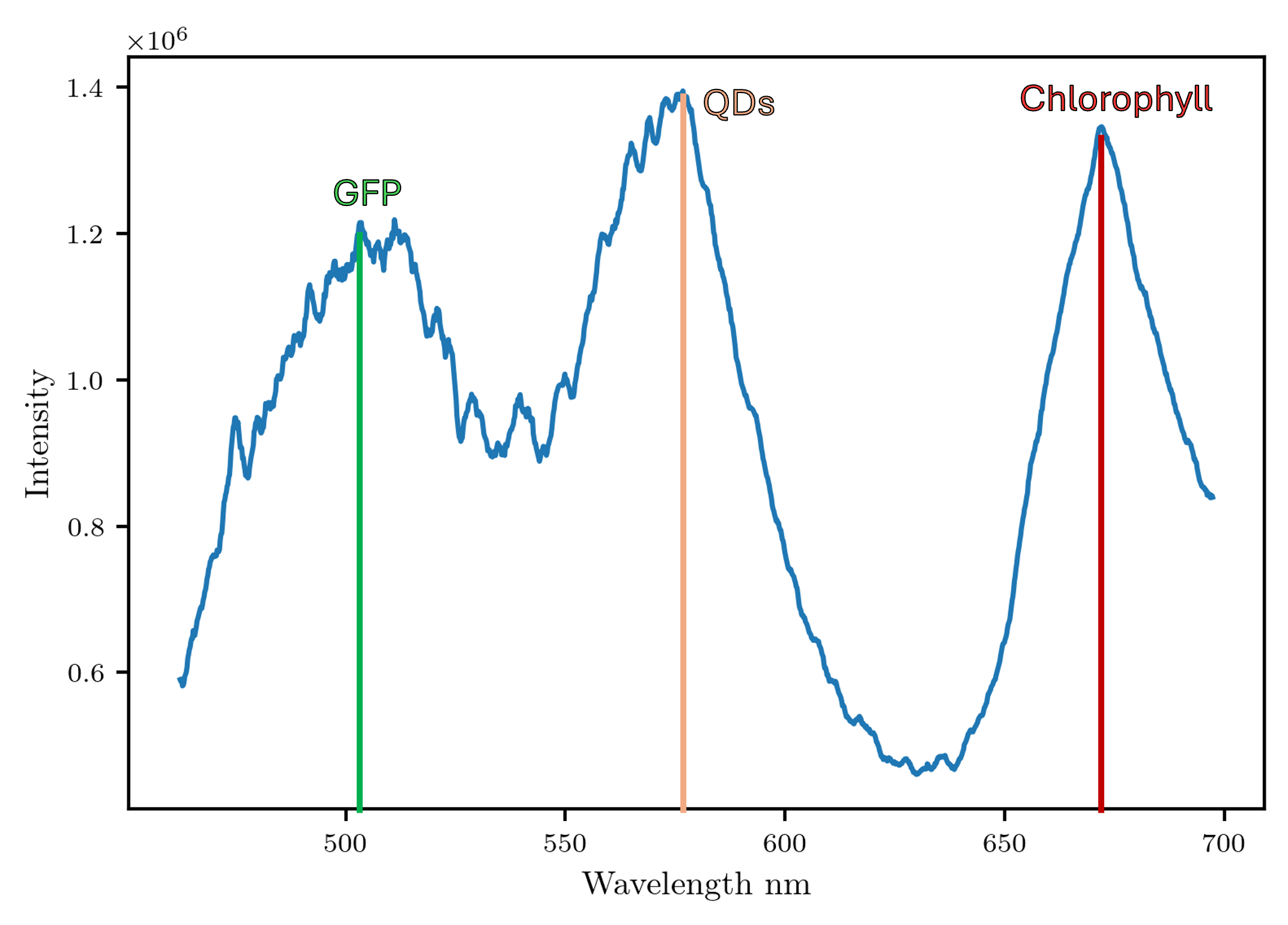} 
\caption{\label{fig:1} Photoluminescence (PL) spectrum of the tobacco leaf showing distinct emission peaks for Green Fluorescent Protein (GFP) at 500 nm, quantum dots (QDs) at 585 nm, and chlorophyll at 680 nm.}
\end{figure}

The successful emission of QDs was confirmed through a photoluminescence spectrum, as illustrated in Figure 6. The plot highlights distinct emission peaks corresponding to the key components introduced into the tobacco leaf cells. Specifically, the GFP exhibited a strong peak around 500 nm, indicating successful expression within the plant cells. The QDs displayed a prominent emission peak at 585 nm, verifying their presence and functionality as quantum light emitters. Additionally, a characteristic chlorophyll fluorescence peak was observed at approximately 680 nm, reflecting the natural autofluorescence of the plant. These well-defined peaks confirm the successful integration and activity of both GFP and QDs within the leaf cells, alongside the expected chlorophyll fluorescence background.

We generated four quantum fingerprints for leaves under LL and HL, as illustrated in Figure 7. The data showed noticeable differences between tobacco plants with low and high-light conditions demonstrating the impact of photosynthetic activity on the quantum properties of the emitted light. The fingerprint is visualized as a 4 × 4 2D plot, where each detector pair correlation contributes 13 data points. These points appear as small rectangles on the plot, resulting in 208 segments across the entire plot. The diagonal elements always show zero as they represent correlations between the same detectors, and the off-diagonal elements represent the time correlation values between different detector pairs. 

\begin{figure}[htbp]
\centering
\includegraphics[width=15.0cm]{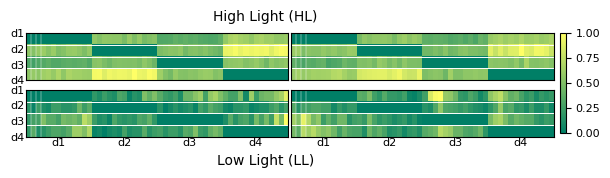} 
\caption{\label{fig:1} Quantum fingerprints for healthy leaf grown in HL conditions (top), and an unhealthy leaf grown in LL conditions (bottom).}
\end{figure} 

These results suggest that quantum readings are affected by differences in photosynthetic productivity and health status. Using a novel quantum-based fingerprinting concept, we developed a higher-order photon correlation to optimize plant productivity. 

The probabilities of the healthy and unhealthy validation leaves are summarized in Table 1. The model predicted a 91\% probability of being healthy and a 9\% probability of being unhealthy for the healthy validation leaf. Similarly, it predicted an 84\% probability of being unhealthy and a 16\% probability of being healthy for the unhealthy validation leaf. This confirms the accurate identification of leaves as healthy or unhealthy using the CNN model.

\begin{table}[h!]
\centering
\resizebox{1.0\textwidth}{!}{ 
\begin{tabular}{|c|c|c|c|}
\hline
\textbf{Validation Type}       & \textbf{Time/Step} & \textbf{Healthy Probability} & \textbf{Unhealthy Probability} \\ \hline
Healthy Validation Leaf        & 0s 54ms/step      & 0.91                         & 0.09                           \\ \hline
Unhealthy Validation Leaf      & 0s 46ms/step      & 0.16                         & 0.84                           \\ \hline
\end{tabular}
}
\caption{Validation Results for Healthy and Unhealthy Leaves}
\end{table}

Ongoing experiments aim to explore further the link between quantum profiles and the optimization of productivity of tobacco plants.

\section{Conclusion}

Our study demonstrates the potential to link quantum data to plant photosynthetic health under varying growth conditions, revealing significant differences between high-light (HL)  and low-light (LL) environments. Light intensity, the critical growth condition we altered, significantly affected several photosynthetic parameters.  The intensity of incoming light was directly related to the amount of light used for photosynthesis, and the ability to dissipate excess light energy as heat. As expected, chlorophyll content was also directly related to light intensity, since chlorophyll is the primary light-capturing molecule in plants and the activity of Photosystem I was more efficient in HL leaves. Photosystem I activity infers that HL leaves are healthier since they have a higher fraction of molecules converting light into chemical energy. Though the photosynthetic parameters we analyzed do not provide an overall measurement of photosynthetic activity for a plant, they establish a baseline indication of plant photosynthetic state. Incorporation of photosynthetic parameters such as stomatal conductance, electron transport rate, and protein and enzyme activity would create a more comprehensive measurement of plant health \cite{Li2021Comparison,Vico2023Photosynthetic}. Correlating the combined, synergistic effects of these parameters with quantum profiles will further train our machine learning algorithm, and correlate plant responses to environmental stressors.

In this study, QDs acting as quantum light emitters, were successfully introduced into tobacco leaves using the biolistic transformation method. This innovative approach allowed for fluorescence measurements, bypassing the limitations imposed by chlorophyll autofluorescence. Time-resolved correlation patterns were analyzed using a CNN model, resulting in the classification of HL and LL leaves and the successful creation of their respective fingerprints. These fingerprints provided unique insights into the quantum fluorescence properties of leaves under different conditions. The CNN model was highly effective in identifying HL and LL leaves by analyzing correlation data and assigning probabilities to each classification. HL leaves were identified with 91\% probability, while LL leaves were identified with 84\% probability. The successful use of machine learning in this context significantly advanced our research, providing a powerful tool for efficient analysis.

Our findings provide a new method for correlating quantum light from quantum dots to photosynthesis in tobacco plants, which offers a novel way for monitoring plant health. Our results encourage further research on species with significant applications in agriculture, due to the heightened frequency and intensity of environmental conditions. As part of our future plan, we aim to refine and expand the approach to analyze a wider variety of plants and algae, including those from different species and under diverse environments (i.e. water-depleted, extreme heat, viral-infected). By advancing the integration of quantum technology and machine learning, this research lays the groundwork for innovative approaches to monitoring and improving plant productivity.

\nocite{*}
\bibliographystyle{plain} 
\bibliography{sample}

\end{document}